\documentclass[runningheads]{llncs}

\RequirePackage[english]{babel}
\usepackage{graphicx,epsfig}

\begin{document}

\title{Nodal two-dimensional solitons in
nonlinear parametric resonance}

\author{N.V. Alexeeva\inst{1} \and
E.V. Zemlyanaya\inst{2} }

\institute{Department of Mathematics, University of Cape Town,
South Africa; \email{nora@figaro.mrh.uct.ac.za} \and Joint
Institute for Nuclear Research, Dubna 141980, Russia;
\email{elena@jinr.ru} }

\titlerunning{Nodal two-dimensional solitons in nonlinear parametric resonance}
\toctitle{Nodal two-dimensional solitons in nonlinear parametric
resonance}

\authorrunning{N.V. Alexeeva and E.V. Zemlyanaya}
\tocauthor{N.V. Alexeeva and E.V. Zemlyanaya}

\maketitle

\begin{abstract}
The parametrically driven damped nonlinear Schr\"odinger equation
serves as an amplitude equation for a variety of resonantly forced
oscillatory systems on the plane. In this note, we consider its
nodal soliton solutions. We show that although the nodal solitons
are stable against radially-symmetric perturbations for
sufficiently large damping coefficients, they are always unstable
to azimuthal perturbations. The corresponding break-up scenarios
are studied using direct numerical simulations. Typically, the
nodal solutions break into symmetric ``necklaces" of stable
nodeless solitons.

\end{abstract}

{\bf 1.} Two-dimensional  localised oscillating structures,
commonly referred to as oscillons, have been detected in
experiments on vertically vibrated layers of granular material
\cite{Swinney}, Newtonian fluids  and suspensions
\cite{Faraday,Astruc}. Numerical simulations  established the
existence of stable oscillons in a variety of pattern-forming
systems, including the Swift-Hohenberg and Ginsburg-Landau
equations, period-doubling maps with continuous spatial coupling,
 semicontinuum theories and hydrodynamic models
\cite{Astruc,numerical}. These simulations provided a great deal
of insight into the phenomenology of the oscillons; however, the
mechanism by which they acquire or loose their stability remained
poorly understood.

In order to elucidate this mechanism, a simple model of a
parametrically forced oscillatory medium was proposed recently
\cite{us}. The model comprises a two-dimensional lattice of
diffusively coupled, vertically vibrated pendula. When driven at
the frequency close to their double natural frequency, the pendula
execute almost synchronous librations whose slowly varying
amplitude satisfies the 2D parametrically driven, damped nonlinear
Schr\"odinger (NLS) equation. The NLS equation was shown to
support radially-symmetric, bell-shaped (i.e. nodeless) solitons
which turned out to be stable for sufficiently large values of the
damping coefficient. These stationary solitons of the amplitude
 equation correspond to the spatio-temporal envelopes of
the oscillons in the original lattice system. By reducing the NLS
to a finite-dimensional system in the vicinity of the soliton, its
stabilisation mechanism (and hence, the oscillon's stabilisation
mechanism) was clarified \cite{us}.

In the present note we consider a more general class of
radially-symmetric solitons of the parametrically driven, damped
NLS on the plane, namely solitons with nodes. We will demonstrate
that these solitons are unstable against azimuthal modes, and
analyse the evolution of this instability.

{\bf 2.} The  parametrically driven, damped NLS equation has the
form:
\begin{equation}
i \psi_t + \nabla^2 \psi+ 2|\psi|^2 \psi - \psi = h \psi^* -i
\gamma \psi. \label{2Dnls}
\end{equation}
Here $\nabla^2=\partial^2 /\partial x^2 +\partial^2 /\partial
y^2.$ Eq.(\ref{2Dnls}) serves as an amplitude equation for a wide
range of nearly-conservative two-dimensional oscillatory systems
under parametric forcing. This equation was also used as a
phenomenological model of nonlinear Faraday resonance in water
\cite{Astruc}. The coefficient $h>0$ plays the role of the
driver's strength and $\gamma>0$ is the damping coefficient.

\begin{figure}[t]
\begin{center}
\psfig{file=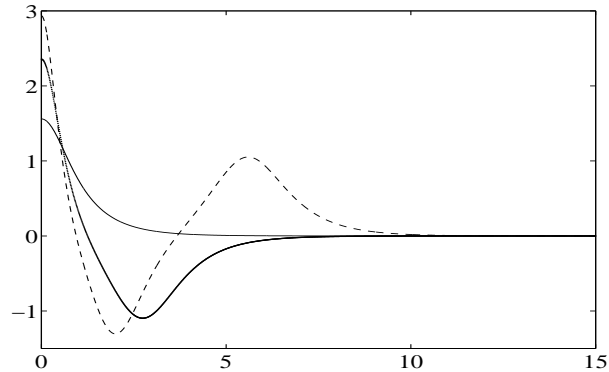,width=8cm,height=5.cm}
\end{center}
\caption{\sf Solutions of eq.(\ref{master}):
 ${\cal
R}_0(r)$ (thin continuous line), ${\cal R}_1(r)$ (thick  line),
${\cal R}_2(r)$ (dashed). } \label{fig0}
\end{figure}

We start with the discussion of its nodeless solitons and their
stability. The exact (though not explicit) stationary
radially-symmetric solution is  given by
\begin{equation}
\psi_0= {\cal A} e^{- i \theta} \, {\cal R}_0({\cal A} r),
\label{soliton}
\end{equation}
where $r^2=x^2+y^2$,
 $${\cal A}^2=1 + \sqrt{h^2-\gamma^2}, \quad
\theta= \frac12 \arcsin\left(\frac{\gamma}{h}\right),$$ and ${\cal
R}_0(r)$ is the bell-shaped nodeless solution of the equation
\begin{equation}
 {\cal R}_{rr} +\frac 1 r {\cal R} - {\cal R} + 2 {\cal R}^3 =0
  \label{master}
\end{equation}
with the boundary conditions ${\cal R}_r(0)={\cal R}(\infty)=0$.
(Below we simply write ${\cal R}$ for ${\cal R}_0$.)
Solutions of eq.(\ref{master}) are well documented in literature
\cite{Rypdal}; see Fig.\ref{fig0}.

\begin{figure}[t]
\begin{center}
\psfig{file=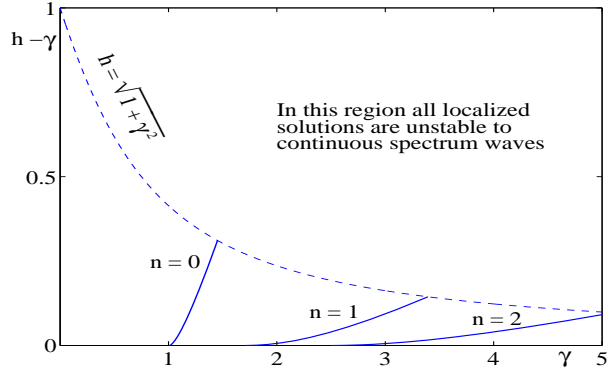,width=8cm,height=5.cm}
\end{center}
\caption{\sf Stability diagram for two-dimensional solitons. The
$(\gamma, h-\gamma)$-plane is used for visual clarity. No
localised or periodic attractors exist for $h< \gamma$ (below the
horisontal axis).  The region of stability of the soliton
$\psi_{0}$  lies  to the right of the
 solid curve marked ``$n=0$". Also shown are the regions of stability of
 the solitons $\psi_1$ and $\psi_2$ with respect to the radially-symmetric
 perturbations. (These lie to the right of the corresponding curves in the
 figure.)
} \label{chart}
\end{figure}

{\bf 3.} To examine the stability of the solution (\ref{soliton})
 with nonzero $h$ and $\gamma$, we
linearise eq.(\ref{2Dnls}) in the small perturbation
$$
\delta \psi({\bf x},t)= e^{(\mu-\Gamma) {\tilde t}-i
\theta_{\pm}}[u({\tilde {\bf x}})+i v({\tilde {\bf x}})],
$$
where ${\tilde {\bf x}}={\cal A} {\bf x}$, ${\tilde t}={\cal A}^2
t$. This yields an eigenvalue problem
\begin{equation}
 L_1 u =- (\mu+ \Gamma )v, \quad
(L_0  - \epsilon)v = (\mu- \Gamma )u, \label{EV_gamma}
\end{equation}
where $\Gamma = \gamma /{\cal A}^2$ and the operators
\begin{equation}
L_0  \equiv - {\tilde \nabla}^2+1 -2 {\cal R}^2({\tilde r}), \quad
L_1 \equiv L_0 -4 {\cal R}^2({\tilde r}), \label{op_nabla}
\end{equation}
with ${\tilde \nabla}^2= \partial^2/\partial {\tilde
x}^2+\partial^2/\partial {\tilde y}^2$. (We are dropping the
tildas below.) For further convenience, we introduce the positive
quantity $\epsilon=  2 \sqrt{h^2-\gamma^2}/{\cal A}^2$.
 Fixing $\epsilon$ defines a curve
on the $(\gamma,h)$-plane:
\begin{equation}
h= \sqrt{ \epsilon^2/(2- \epsilon)^2 + \gamma^2}.
\label{h_of_gamma}
\end{equation}
Introducing
\begin{equation}
\lambda^2=\mu^2-\Gamma^2 \label{mugamma}
\end{equation}
and performing the transformation \cite{BBK} $$v({\bf x}) \to (\mu
+\Gamma) \lambda^{-1} v({\bf x}),$$ reduces eq.(\ref{EV_gamma})
to a {\it one-}parameter eigenvalue problem:
\begin{equation}
(L_0-\epsilon) v= \lambda u, \quad L_1 u= -\lambda v. \label{EV}
\end{equation}

We first consider the stability with respect to radially symmetric
perturbations $u=u(r), \, v=v(r)$. In this case the operators
(\ref{op_nabla}) become
\begin{equation}
L_0  =- \frac{d^2}{dr^2} -\frac{1}{r} \frac{d}{dr}+1 -2 {\cal
R}^2(r), \quad L_1 = L_0 -4 {\cal R}^2(r). \label{op_radsym}
\end{equation}
In the absence of the damping and driving, all localised initial
conditions in the unperturbed 2D NLS equation are known to either
disperse or blow-up in finite time \cite{Rypdal,Malkin,collapse}.
It turned out, however, that the soliton $\psi_0$ stabilises as
the damping $\gamma$ is increased above a certain value \cite{us}.
The stability condition
 is
$\gamma \ge \gamma_c$, where
 \begin{equation}
\gamma_c=\gamma_c(\epsilon) \equiv \frac{2}{2-\epsilon} \cdot
\frac{{\rm Re\/} \lambda(\epsilon)
  \, {\rm Im\/} \lambda(\epsilon)}
 {\sqrt{({\rm Im\/} \lambda)^2- ({\rm Re\/} \lambda)^2}}.
 \label{gce}
 \end{equation}
We obtained $\lambda(\epsilon)$ by solving the eigenvalue problem
(\ref{EV}) directly.   Expressing $\epsilon$ via $\gamma_c$ from
(\ref{gce}) and feeding into (\ref{h_of_gamma}), we get the
stability boundary on the $(\gamma,h)$-plane (Fig.\ref{chart}).

{\bf 4.} To study the stability to asymmetric perturbations we
factorise, in (\ref{EV}),
$$u({\bf x})={\tilde u}(r) e^{i m \varphi}, \quad v({\bf x})={\tilde v}(r)
e^{i m \varphi},$$ where $\tan \varphi= y/x$ and $m$ is an
integer. The functions ${\tilde u}(r)$ and ${\tilde v}(r)$ satisfy
the  eigenproblem (\ref{EV}) where the operators (\ref{op_radsym})
should be replaced by
\begin{equation}
L_0^{(m)}
 \equiv L_0
  +{m^2}/{r^2},
\quad L_1^{(m)} \equiv L_1+ {m^2}/{r^2}, \label{m2r2}
\end{equation}
respectively. This modified eigenvalue problem can be analysed  in
a similar way to eqs.(\ref{EV}). It is not difficult to
demonstrate  that all discrete eigenvalues of (\ref{EV}) (if any
exist) have to be pure imaginary in this case, and hence the
azimuthal perturbations do not lead to any instabilities of the
solution in question \cite{us}.

\begin{figure}[t]
\begin{center}
\psfig{file=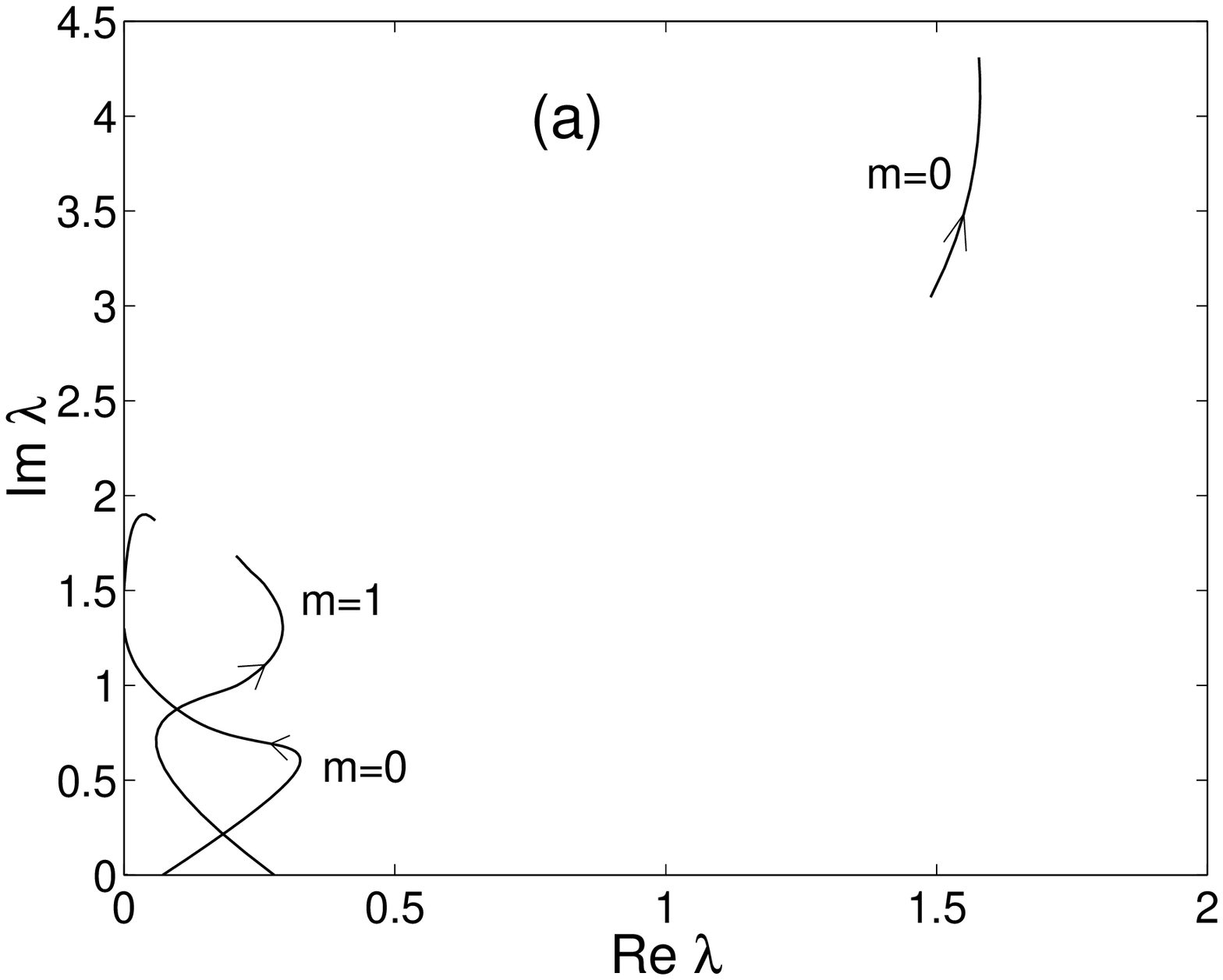,width=6cm,height=4.7cm}\\
\psfig{file=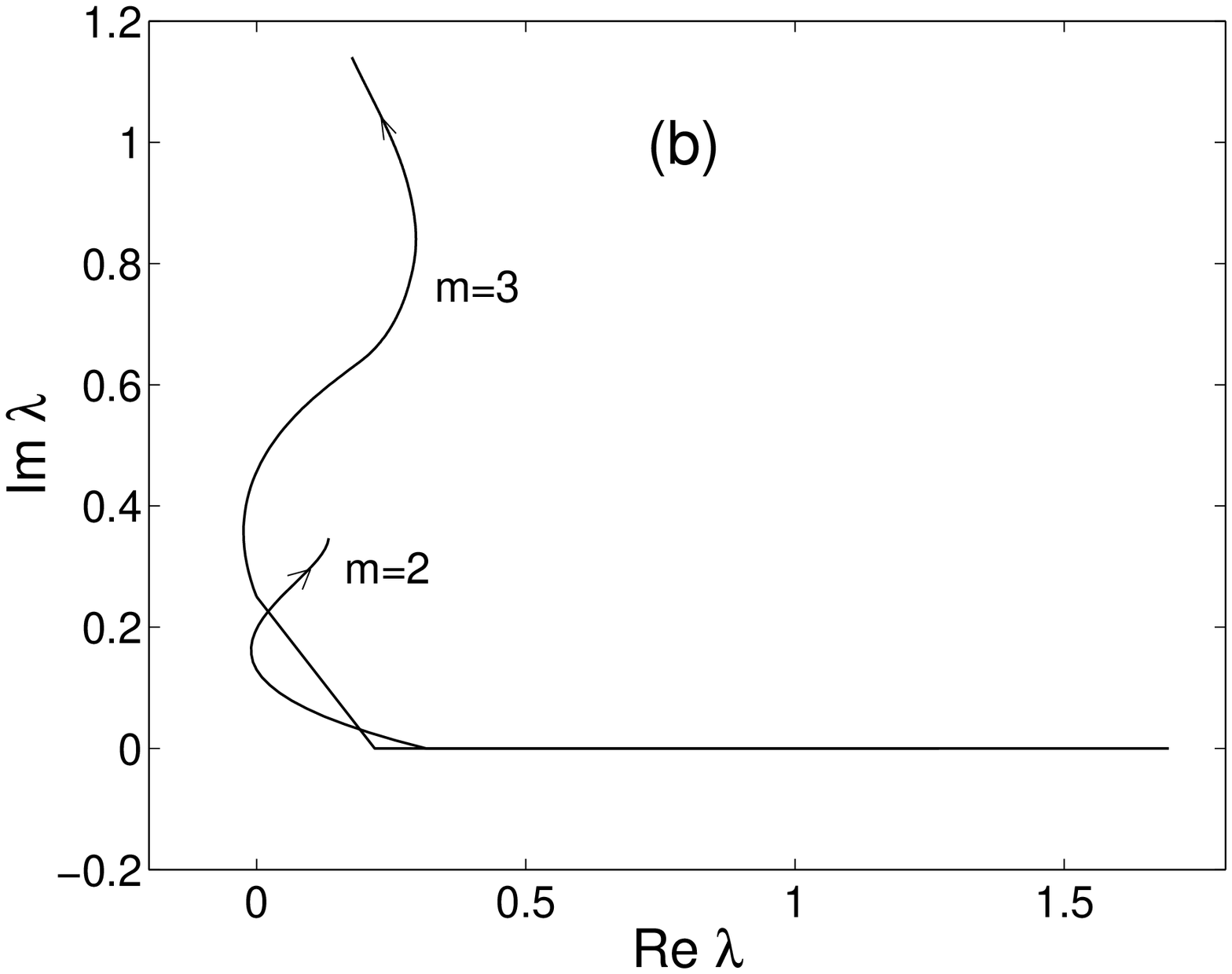,width=6cm,height=4.7cm}
\psfig{file=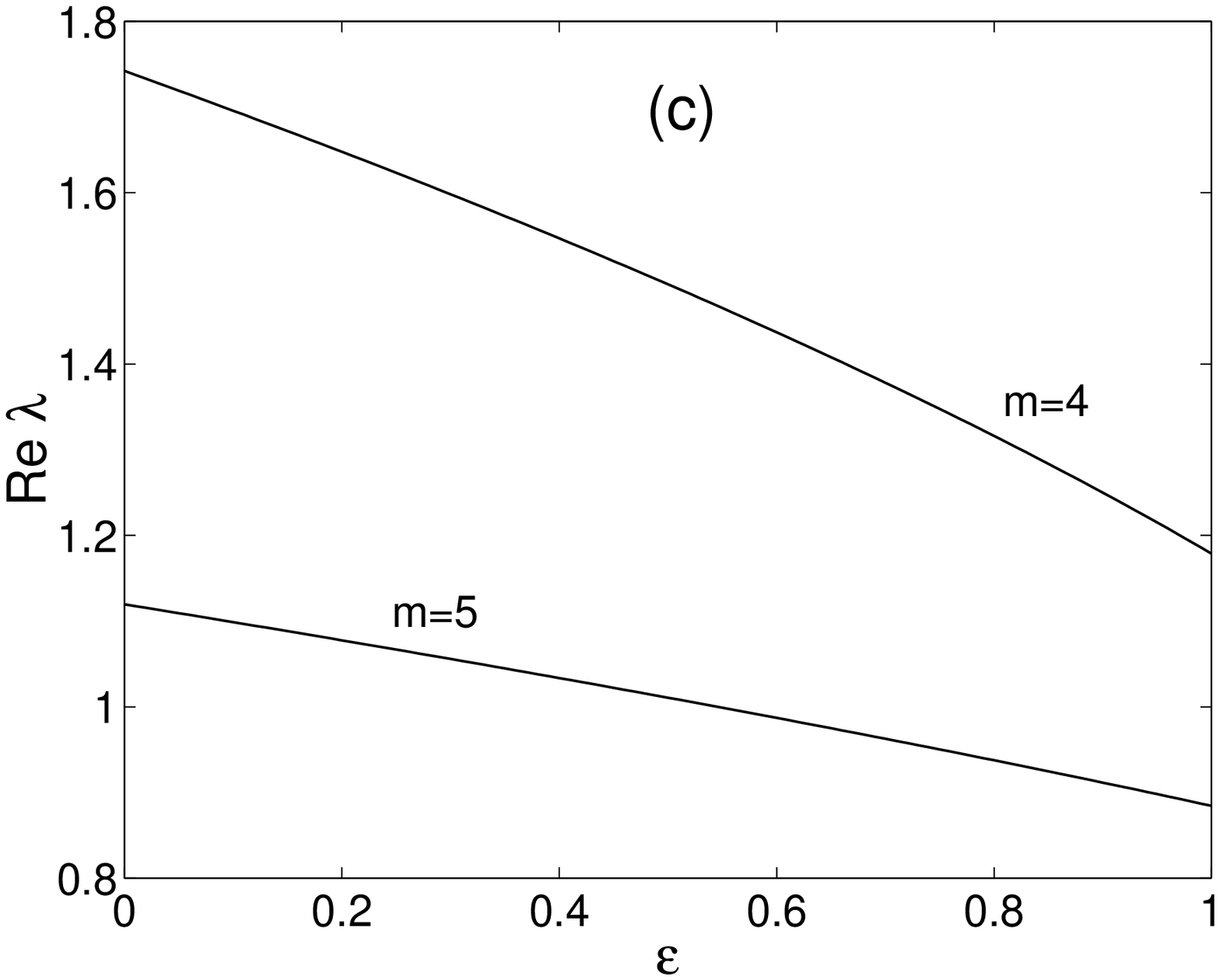,width=6cm,height=4.7cm}
\end{center}
\caption{\sf The discrete eigenvalues  of the linearised operator
(\ref{EV}) for the one-node soliton, $\psi_1$. Panels (a) and (b)
show the {\it complex\/} eigenvalues, $\mbox{Im} \, \lambda$ vs
$\mbox{Re} \, \lambda$. Arrows indicate the direction of increase
of $\epsilon$.
 Panel (c) shows the {\it real\/} eigenvalues,
as functions of $\epsilon$.} \label{fig2}
\end{figure}

\begin{figure}[t]
\begin{center}
\psfig{file=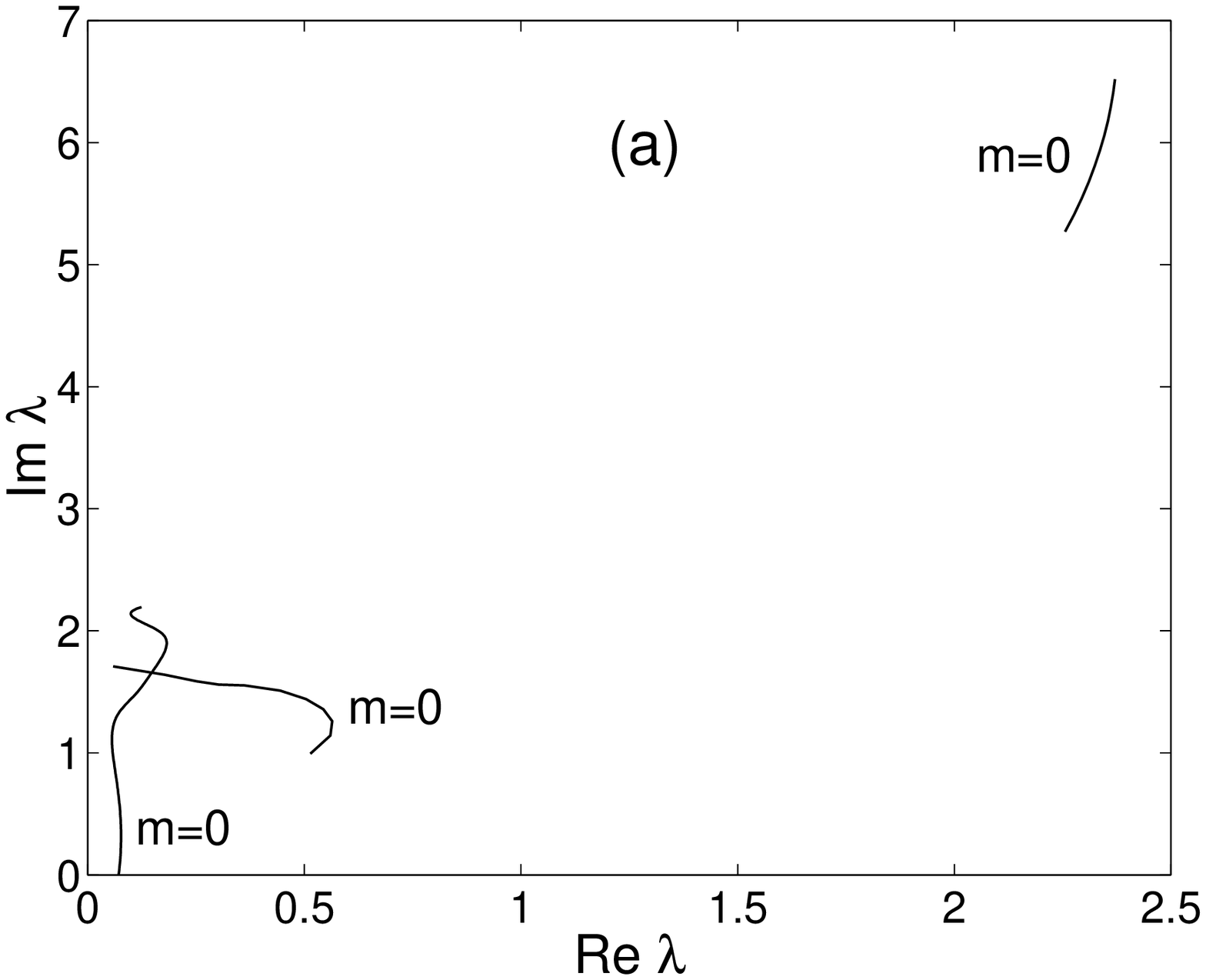,width=6cm,height=4.7cm}
\psfig{file=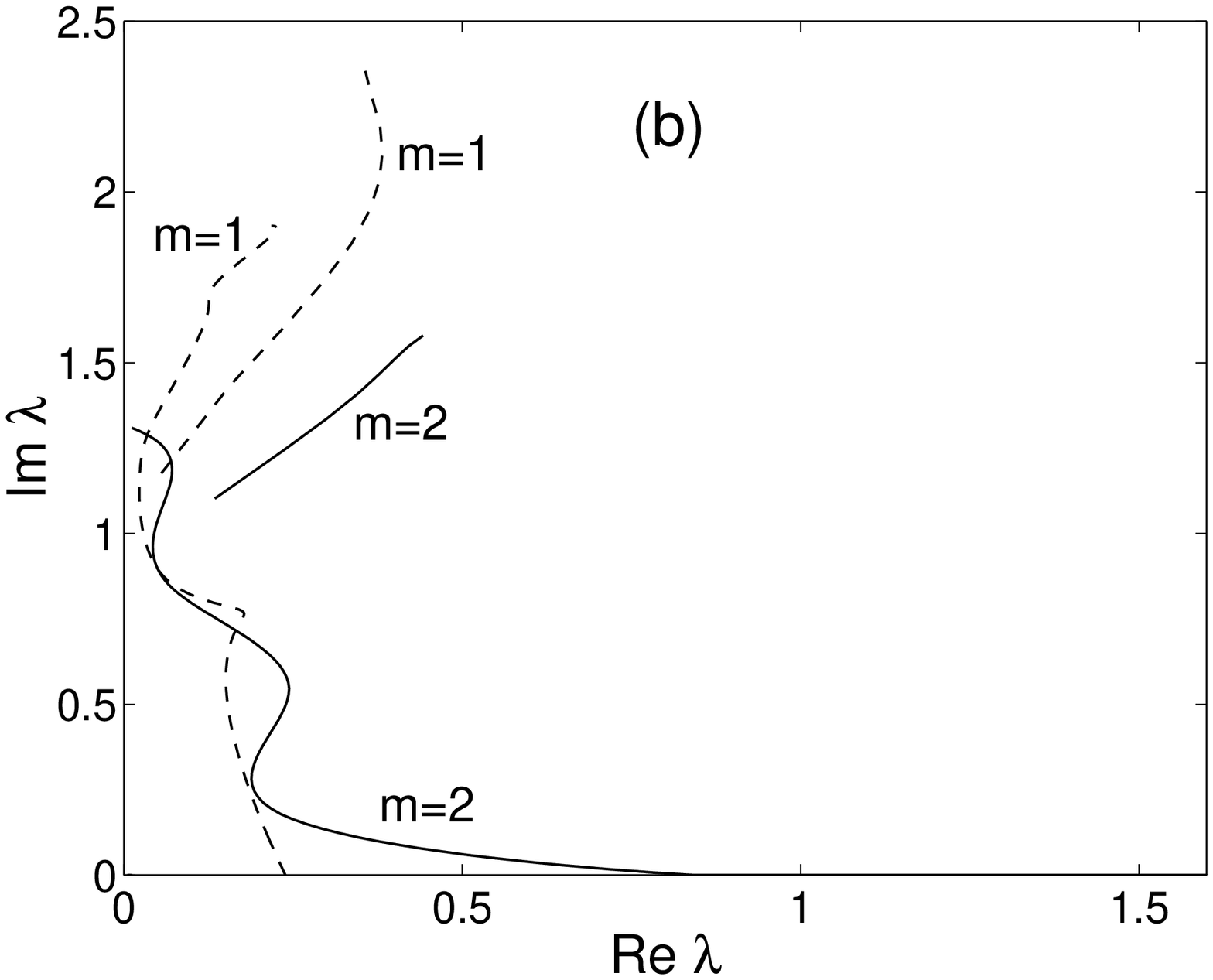,width=6cm,height=4.7cm}\\
\psfig{file=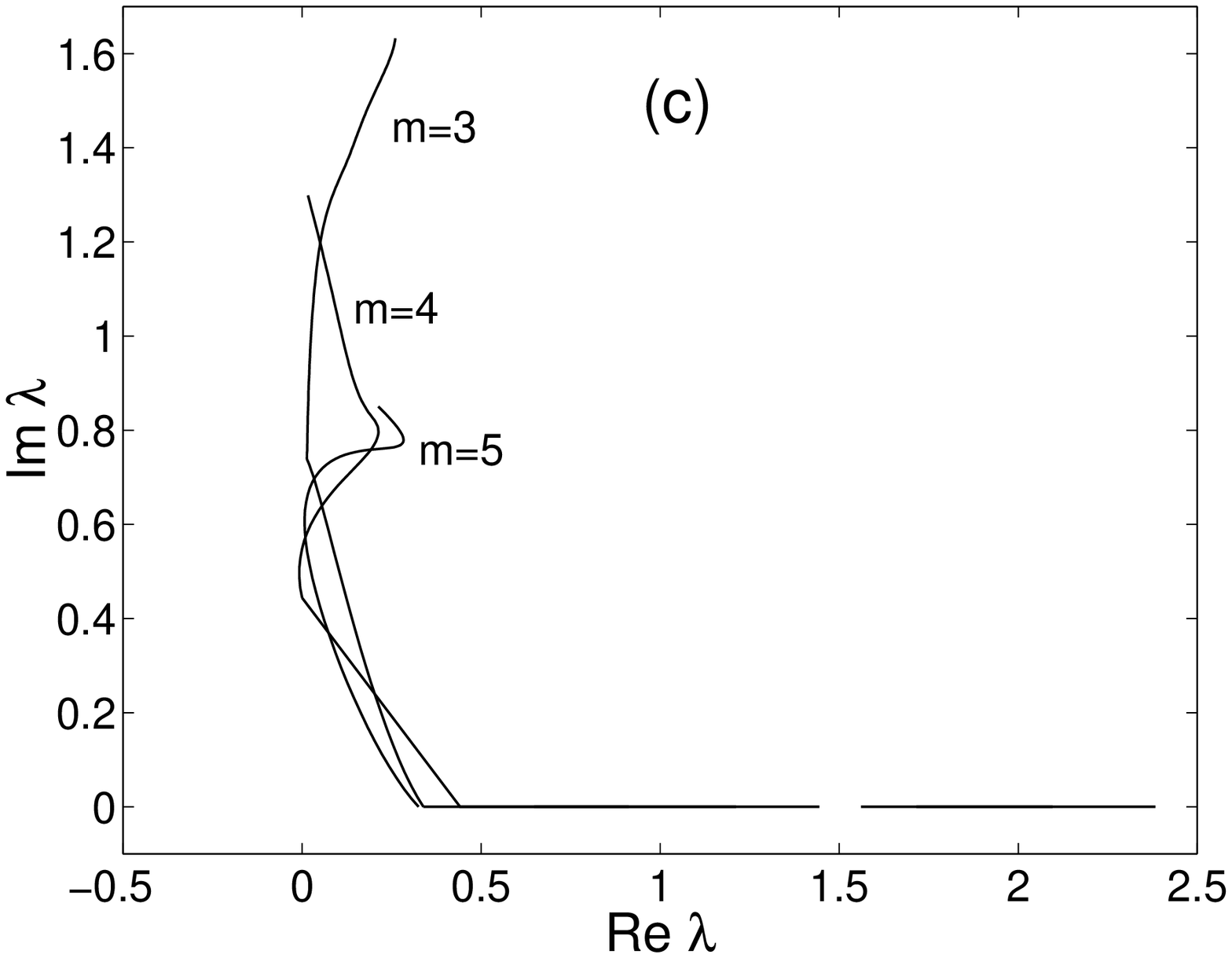,width=6cm,height=4.7cm}
\psfig{file=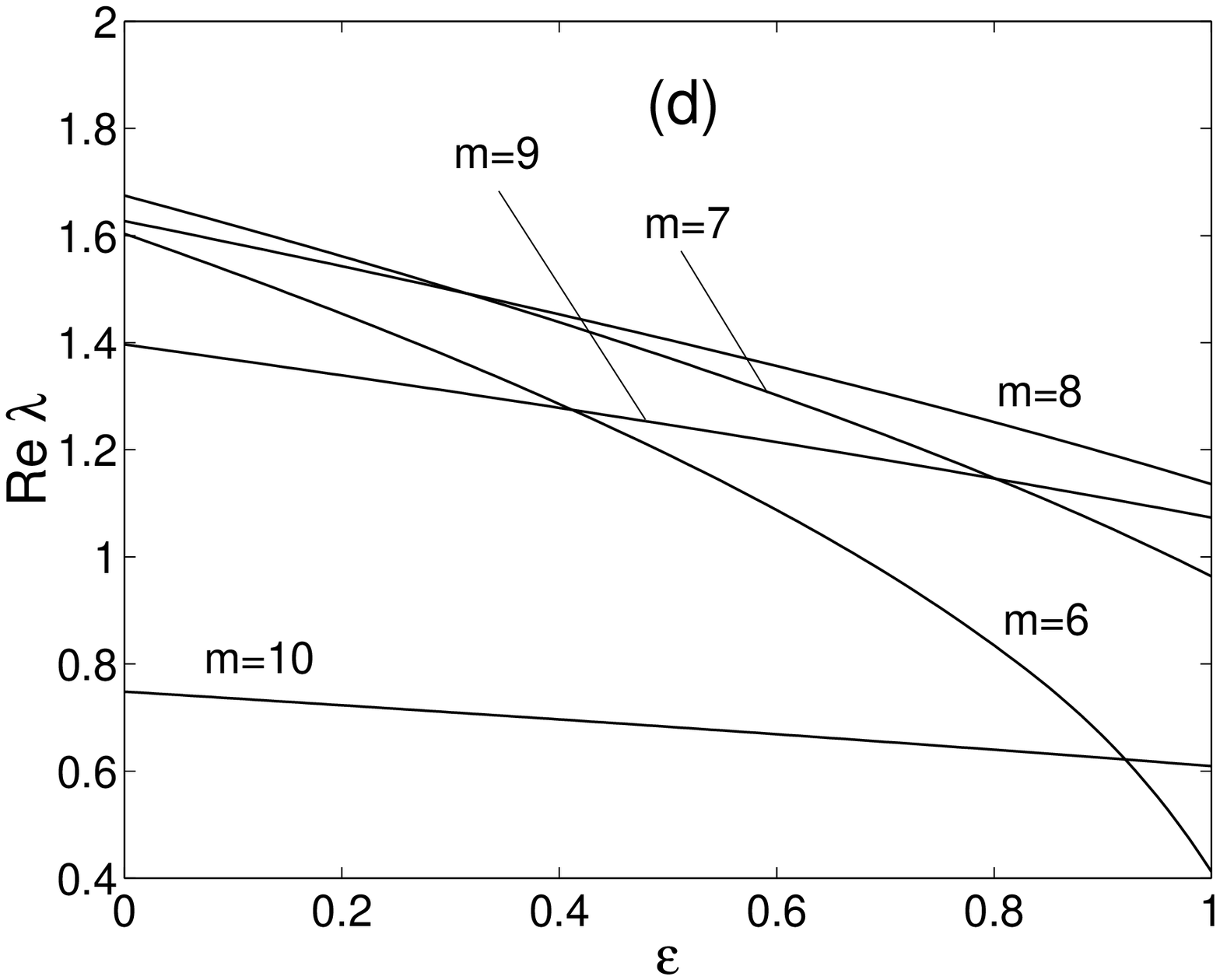,width=6cm,height=4.7cm}\end{center}
\caption{\sf The discrete eigenvalues  of the linearised operator
(\ref{EV}) for the two-node soliton, $\psi_2$. Panels (a),(b), and
(c) show the complex eigenvalues, $\mbox{Im} \, \lambda$ vs
$\mbox{Re} \, \lambda$. Panel (d) shows the real eigenvalues, as
functions of $\epsilon$.} \label{fig3}
\end{figure}

\begin{figure}[t]
\begin{center}
\psfig{file=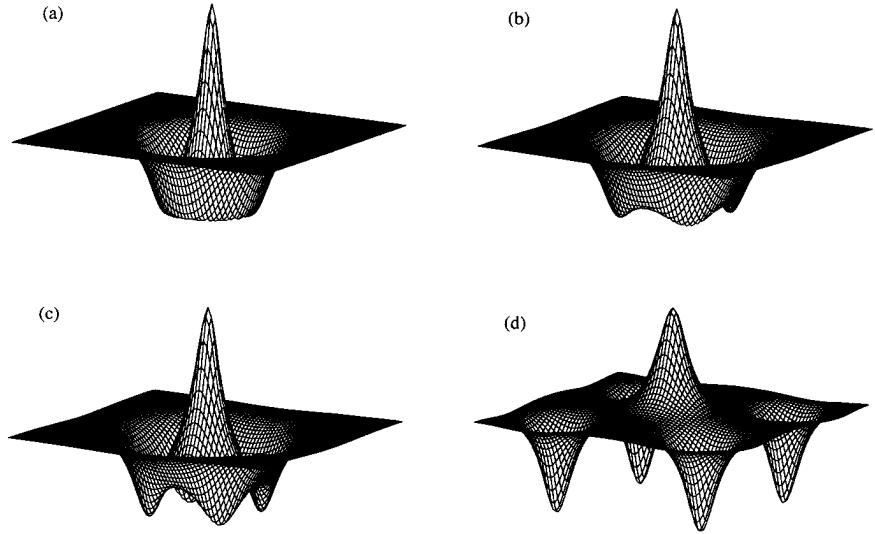}
\end{center}
\caption{\sf Evolution of the azimuthal instability of the
one-node soliton. (a): the initial condition, soliton $\psi_1$ ;
(b) and (c):
 dissociation of the ring-like ``valley'' into 4 nodeless solitons;
 (d): divergence of the fragments.
 Here $\gamma=3.5$ and $h=3.6$;  shown is $\mbox{Re} \, \psi$.
 Note the change of the
vertical scale in (d).} \label{fig4}
\end{figure}


\begin{figure}
\begin{center}
\psfig{file=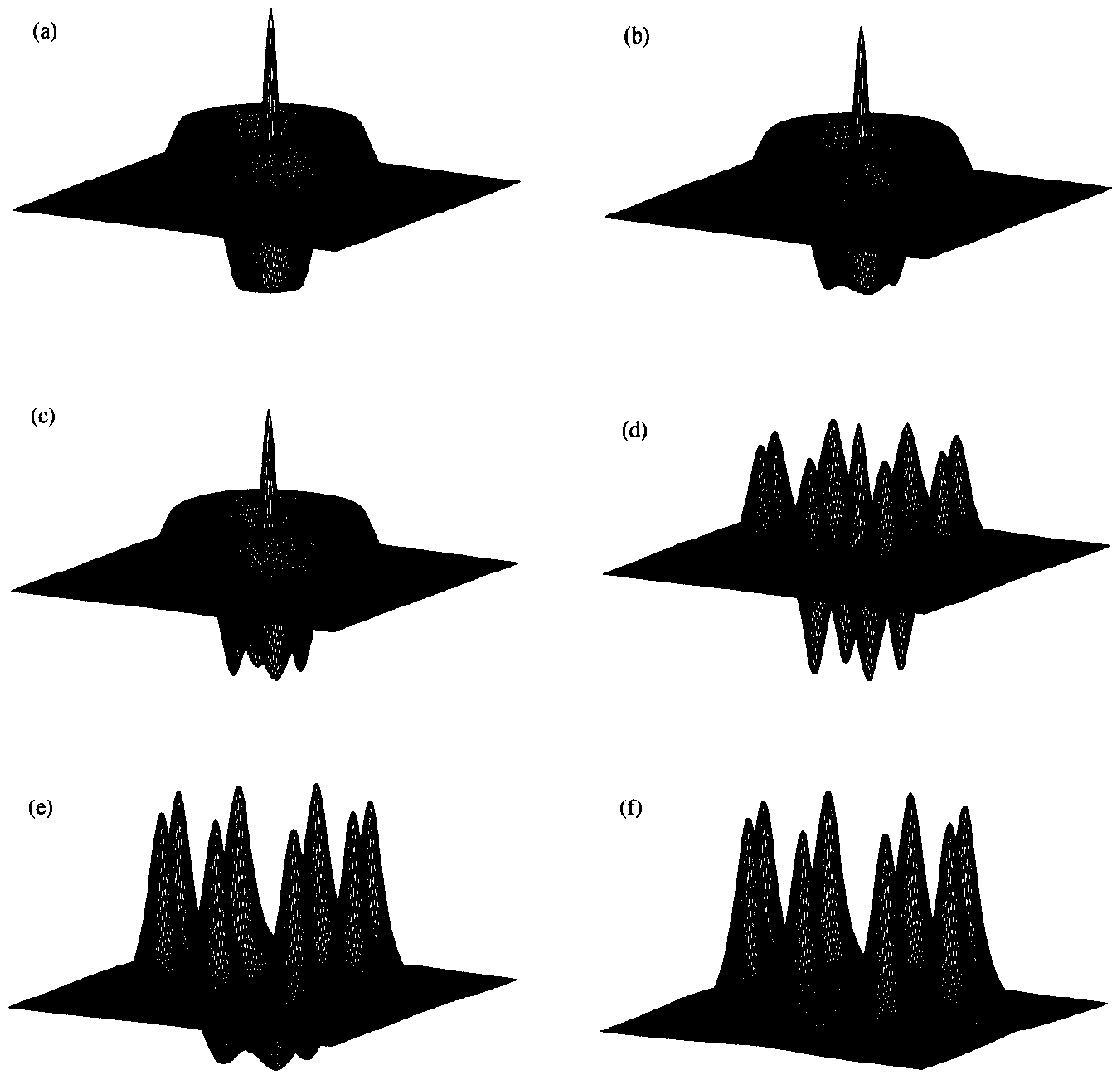}
\end{center}
\caption{\sf The evolution of the azimuthal instability of the
two-node soliton $\psi_2$. (a): the initial condition; (b)-(c):
the rapid  dissociation of the ``valley" into 4 nodeless solitons;
(c)-(d): a slower decay of the ``ridge" into 8 solitons $\psi_0$;
(e)-(f): the annihilation of the internal ring and the central
soliton, and the repulsion of the persisting 8 solitons. Here
$\gamma=4.5$ and $h=4.53$; shown is $\mbox{Re} \, \psi$. Note the
change of the vertical scale in (e)-(f) w.r.t. that in (a)-(d).}
\label{fig5}
\end{figure}

{\bf 5.} Besides the nodeless solution ${\cal R}_0(r)$, the
``master" equation (\ref{master}) has solutions ${\cal R}_n(r)$
with $n$ nodes, $n \ge 1$. (See Fig.\ref{fig0}). These give rise
to a sequence of nodal solutions $\psi_n$ of the damped-driven NLS
(\ref{2Dnls}), defined by eq.(\ref{soliton}) with ${\cal R}_0 \to
{\cal R}_n$. To examine the stability of the
$\psi_{n}$, 
we  solved the eigenvalue problem (\ref{EV}) numerically,
with 
 operators $L_{0,1}^{(m)}$ as in
(\ref{m2r2}). The {\it radial\/} stability properties of the nodal
solitons turned out to be similar to those of the nodeless soliton
$\psi_0$. Namely, the $\psi_n$ solutions are stable against
radially-symmetric perturbations  for sufficiently large $\gamma$.
The corresponding stability regions for $\psi_1$ and $\psi_2$ are
depicted in Fig.(\ref{chart}). However, the {\it azimuthal\/}
stability properties of the nodal solitons have turned out to be
quite different.

Both $\psi_1$ and $\psi_2$ solutions do have  eigenvalues
$\lambda$ with nonzero real parts  for
 orbital numbers $m \ge 1$.
(See Fig.\ref{fig2} and \ref{fig3}.) Having found  eigenvalues
$\lambda$  for each $\epsilon$, one still has to identify those
giving rise to the largest growth rates
\begin{equation}
\nu= \mbox{Re} \, \mu -\Gamma \label{nu}
\end{equation}
for each pair $(\epsilon, \gamma)$ [or, equivalently, for each
$(h,\gamma)$]. In (\ref{nu}), $\mu$ is reconstructed using
eq.(\ref{mugamma}). The selection of {\it real\/} eigenvalues is
straightforward; in this case we
have the following two simple rules:\\
$\bullet$ If, for some $\epsilon$, there are eigenvalues
$\lambda_1>0$, $\lambda_2>0$ such that $\lambda_1>\lambda_2$, then
$\nu_1>\nu_2$ for this $\epsilon$ and all $\gamma$. That is, of
all real eigenvalues $\lambda$ one  has to
consider only the  largest one.\\
$\bullet$ If, for some $\epsilon$, there is a real eigenvalue
$\lambda_1>0$ and a complex eigenvalue $\lambda_2$, with
$\mbox{Re}\, \lambda_2>0$ and $\lambda_1 > \mbox{Re} \,
\lambda_2$, then $\nu_1>\nu_2$ for this $\epsilon$ and all
$\gamma$. That is, one can ignore all complex eigenvalues with
real parts smaller than a real eigenvalue --- if there is one.

The comparison of two complex eigenvalues is not so
straightforward. In particular, the fact that $\mbox{Re} \,
\lambda_1
> \mbox{Re} \, \lambda_2$ does not necessarily imply that
$\nu_1>\nu_2$. Which  of the two growth rates, $\nu_1$ or $\nu_2$,
is larger will depend on the imaginary parts of $\lambda$, as well
as on $\gamma$.

In figures \ref{fig2} and \ref{fig3}, we illustrate the real and
imaginary parts of the eigenvalues, arising for different $m$, for
the solitons $\psi_1$ and $\psi_2$. The soliton $\psi_1$ has
discrete eigenvalues $\lambda$ associated with orbital numbers
$m=0,1,...,5$ and the soliton $\psi_2$ with
 $m=0,1,...,10$.

\begin{table}
\begin{center}
\begin{tabular}{|c|c|c|c|c|c|c|c|c|} \cline{1-4} \cline{6-9}
m & $\nu$ & Re$\lambda$ & Im$\lambda$&$\qquad$&  m & $\nu$ & Re$\lambda$ & Im$\lambda$ \\
\cline{1-4} \cline{6-9}
         0 & -0.1620 &   1.5797   & 4.2181  &&   0 &   -0.3361  &  2.3531 &   6.1585 \\
         0 & -1.4827 &   0.0609   & 1.8743 && 0 &   -0.5877  &  0.1819 &
         1.7847 \\
                  & & & &&   0 &  -0.4168   &  0.3093 &   1.5572 \\
\cline{1-4} \cline{6-9}
    1 &  -0.8255    & 0.2272    & 1.6198  && 1 &   -0.8089 &   0.3818  &  2.1021 \\
    1 &    2.79e-6   & 0.0033    & 0.0000 &&   1 &   -0.4891 &   0.1111  &  1.6352 \\
    & & & & &  1 &   1.07e-5 & 0.0079 &   0.0000 \\
\cline{1-4} \cline{6-9}
    2 &   -0.3012   & 0.2213    & 1.0602  &&   2 &   -0.3497 &   0.3737  &  1.4597 \\
& & & & & 2 &   -0.0328 &   0.2602  &  0.5128 \\
\cline{1-4} \cline{6-9}
    3 &    0.0872   & 0.5821    & 0.0000 && 3 &    0.5406 &   1.8686  &  0.0000 \\
\cline{1-4} \cline{6-9}
    4 &    0.3689   & 1.2399    & 0.0000 &&   4 &    0.5286 &   1.8462  &  0.0000 \\
\cline{1-4} \cline{6-9}
    5 &    0.2057   & 0.9076    & 0.0000 &&    5 &    0.0263 &   0.3958  &  0.0000 \\
\cline{1-4} \cline{6-9}
&&&&&    6 &    0.1611 &   0.9898  &  0.0000 \\
 \cline{6-9}
 &&&&&    7 &    0.2490 &   1.2392  &  0.0000  \\
 \cline{6-9}
 &&&&&    8 &    0.2783 &   1.3133  &  0.0000 \\
 \cline{6-9}
 &&&&&    9 &    0.2288 &   1.1861  &  0.0000  \\
 \cline{6-9}
 &&&&&   10 &    0.0720 &   0.6567  &  0.0000 \\
\cline{1-4}  \cline{6-9}
   \end{tabular}
   \end{center}
   \caption{\sf Eigenvalues $\lambda$
   and corresponding growth rates
$\nu$ for the
 solitons $\psi_1$ (left panel) and $\psi_2$ (right panel).
   We included only the eigenvalues which can,
 potentially, give rise to the largest growth rate in each
 ``symmetry class" $m$. Some other eigenvalues have been
 filtered out using the above selection rules.}
\end{table}

In order to compare the conclusions based on the linearised
analysis with direct numerical simulations of the unstable
solitons $\psi_1$ and $\psi_2$, we fix some $h$ and $\gamma$ and
identify the eigenvalue with the maximum growth rate in each case.
In the case of the soliton $\psi_1$, we choose $\gamma=3.5$ and
$h=3.6$; the corresponding $\epsilon=0.9146$. The real and
imaginary parts of $\lambda$ for each $m$ as well as the resulting
growth rates $\nu$ are given in Table 1 (left panel). The
eigenvalue with the largest ${\rm Re} \lambda$ is associated with
$m=0$; however, for the given $\epsilon$ and $\gamma$ the
resulting $\nu <0$. (This is because we have chosen a
 point in the ``radially stable'' part of the $(\gamma, h)$-plane,
to the right of the ``$n=1$'' curve in Fig.{\ref{chart}}.) On the
contrary, the growth rates corresponding to the real eigenvalues
associated with
 $m=3,4,5$ are
positive for all $\gamma$. The maximum growth rate is associated
with $m=4$. The corresponding eigenfunctions $u(r)$ and $v(r)$
have a single maximum near the position of the minimum of the
function ${\cal R}_1(r)$; that is, the perturbation is
concentrated near the circular ``valley" in the relief of
$\psi(x,y)|^2$.
 This
observation suggests that for $\gamma =3.4$ and $h=3.5$, the
soliton $\psi_1$ should break into a symmetric pattern of 5
solitons $\psi_0$: one at the origin and four around it.

Next, in the case of the soliton $\psi_2$ we fix $\gamma=4.5$ and
$h=4.53$; this gives $\epsilon=0.6846$. The corresponding
eigenvalues, for each $m$, are presented in Table 1 (right panel).
Again, the eigenvalue with the largest $\mbox{Re} \lambda$ is the
one for $m=0$ but  the resulting $\nu$ is negative. The largest
growth rates ($\nu_3=0.54$ and $\nu_4=0.53$, respectively) are
those pertaining to $m=3$ and $m=4$. The corresponding
eigenfunctions have their maxima near the position of the minimum
of the function ${\cal R}_2(r)$. Therefore, the circular ``valley"
of the soliton $\psi_2$ is expected to break into three or four
nodeless solitons $\psi_0$. (Since $\nu_3$ is so close to $\nu_4$,
the actual number of resulting solitons
--- three or four --- will be very sensitive
to the choice of the initial perturbation.) Next, eigenfunctions
pertaining to $m=5,6,...10$ have their maxima near the second,
lateral, maximum of the function ${\cal R}_2(r)$. The largest
growth rate in this group of eigenvalues arises for $m=8$. Hence,
the circular ``ridge" of the soliton $\psi_2$ should break into 8
nodeless solitons, with this process taking longer than the
bunching of the ``valley" into the ``internal ring" of solitons.

The direct numerical simulations corroborate the above scenarios.
The $\psi_1$ soliton with $\gamma=3.5$ and $h=3.6$ splits into a
constellation of 5 nodeless solitons: one at the origin and  four
solitons of opposite polarity at the vertices of the square
centered at the origin. The emerging nodeless solitons are stable
but  repelling each other, see Fig.\ref{fig4}. Hence, no
stationary nonsymmetric configurations are possible; the
peripheral solitons escape to infinity. The $\psi_2$ soliton with
$\gamma=4.5$ and $h=4.53$ has a more complicated evolution. As
predicted by the linear stability analysis, it dissociates into 13
nodeless solitons: one at the origin, four solitons of opposite
polarity forming a square around it and  eight more solitons (of
the same polarity as the one at the origin) forming an outer ring.
(The fact that the inner ring consists of four and not three
solitons, is due to the square symmetry of our domain of
integration which favours perturbations with $m=4$ over those with
$m=3$.) In the subsequent evolution the central soliton and the
nearest four annihilate
 and only the eight outer solitons persist. They
repel each other and eventually escape to infinity, see
Fig.\ref{fig5}.

In conclusion, our analysis suggests the interpretation of  the
nodal solutions as degenerate, unstable coaxial complexes of the
nodeless solitons $\psi_0$.

{\bf Acknowledgements.} This project was supported by the NRF of
South Africa under grant  2053723. The work of E.Z.    was
supported by   an RFBR grant 03-01-00657.

\end{document}